\def\year{2023}\relax
\documentclass[letterpaper]{article} 
\usepackage{aaai23} 
\usepackage{times} 
\usepackage{helvet} 
\usepackage{courier} 
\usepackage[hyphens]{url} 
\usepackage{graphicx} 
\usepackage{natbib} 
\usepackage{caption} 
\usepackage{type1cm} 
\usepackage{graphicx} 
\usepackage{xspace} 
\usepackage{booktabs} 
\usepackage{multirow} 
\usepackage{bold-extra} 
\usepackage{siunitx} 
\usepackage[vlined,linesnumbered,ruled,noend]{algorithm2e} 
\usepackage{microtype} 
\usepackage{xfrac} 
\usepackage{mathtools} 
\usepackage[capitalise]{cleveref}
\usepackage{subfig}
\usepackage[dvipsnames]{xcolor}
\graphicspath{{../img/}}

\usepackage{marvosym} 

\providecommand{\keywords}[1]
{
 \small	
 \textbf{\textit{Keywords---}} #1
}

\title{Beyond the Headlines: Understanding Sentiments and Morals\\ Impacting Female Employment in Spain}

\author {
 Oscar Araque\textsuperscript{\rm 1}, 
 Luca Barbaglia\textsuperscript{\rm 2}, 
 Francesco Berlingieri\textsuperscript{\rm 2}, 
 Marco Colagrossi\textsuperscript{\rm 2}, 
 Sergio Consoli\textsuperscript{\rm 2}\thanks{Corresponding author. \Letter~sergio.consoli@ec.europa.eu}, 
 Lorenzo Gatti\textsuperscript{\rm 3}, 
 Caterina Mauri\textsuperscript{\rm 2}, 
Kyriaki Kalimeri\textsuperscript{\rm 4}
}
\affiliations {
 \textsuperscript{\rm 1}Universidad Politécnica de Madrid, ETSI Telecomunicación, Madrid, Spain\\
 \textsuperscript{\rm 2} European Commission, Joint Research Centre (DG JRC), Ispra (VA), Italy\\ 
 \textsuperscript{\rm 3} University of Twente, Enschede, The Netherlands\\
 \textsuperscript{\rm 4} ISI Foundation, Turin, Italy\\
}

\begin{document}

\maketitle

\begin{abstract}
After decades of improvements in the employment conditions of females in Spain, this process came to a sudden stop with the Great Spanish Recession of 2008.
In this contribution, we analyse a large longitudinal corpus of national and regional news outlets employing advanced Natural Language Processing techniques to capture the valence of mentions of gender inequality expressed in the Spanish press.
The automatic analysis of the news articles does indeed capture the known hardships faced by females in the Spanish labour market.
Our approach can be straightforwardly generalised to other topics of interest.
Assessing the sentiment and moral values expressed in the articles, we notice that females are, in the majority of cases, concerned 
more than males when there is a deterioration in the overall labour market conditions, based on newspaper articles. This behaviour has been present in the entire period of study (2000--2022) and looked particularly pronounced during the economic crisis of 2008 and the recent COVID-19 pandemic.
Most of the time, this phenomenon looks to be more pronounced at the regional level, 
perhaps caused by a significant focus on local labour markets rather than on aggregate statistics or because, in local contexts, females might suffer more from an isolation or discrimination condition. 
Our findings contribute to a deeper understanding of the gender inequalities in Spain using alternative data, informing policymakers and stakeholders.
\end{abstract}


\section{Introduction}

Spain has experienced significant changes in its labour market over the past few decades \cite{RePEc:fda:fdaeee:eee2016-10}\footnote{\url{https://www.rabobank.com/knowledge/d011323133-the-crisis-and-recovery-of-spains-labour-market}}. Despite some general progress, gender inequalities in the labour market persist \cite{iloreport}: females continue to face significant challenges when it comes to accessing employment opportunities and achieving wage parity with their male counterparts \cite{genderUnemployment}\footnote{
Here we refer to ``gender'' as the socially constructed roles, behaviours, activities and attributes that a given society considers appropriate for women, men, or other gender identities \cite{CouncilofEuropeConvention2011}.}. 
This is due to a range of social, cultural, and economic factors, including traditional gender roles, discriminatory hiring practices, and inadequate policies to support female participation in the workforce.
The impact of gender inequalities in the labour market extends beyond individual female and affects the wider society, including the economy, social stability, and human rights \cite{genderUnemployment,iloreport}.

The role of news is critical in raising awareness, playing a vital role in shaping public opinion and influencing policy decisions. 
In this study, we analyse a large corpus of news covering the leading national and regional outlets of the Spanish press over 20 years. 
Our goal is to quantitatively assess the views expressed in these outlets insofar as they relate to labour market outcomes for females.

We apply Natural Language Processing (NLP) techniques to assess the evolution of news narratives over time and geographical location. 
Our framework is based on the fine-grained aspect-based approach by \citet*{Consoli2022a} for sentiment analysis (SA):
building on this reference, we dive deeper into the moral narratives around the unemployment topic in this work, analysing the moral values surrounding the discourse. 
SA has been widely used to enrich and sharpen economic analysis \cite{Consoli20211,Einav2014} as well as to forecast economic and financial variables \cite[e.g.,][]{Consoli2021,Consoli2022,Barbaglia2022}. It has also been extensively used elsewhere in the social sciences, especially in political science \cite[e.g.,][]{Tumasjan2011402,Burnap2016230,Mohammad2015480,Iyyer20141113,Hofmann2013387,Xu2012656,Ceron2014340}.
Our approach can successfully analyse relevant information from textual sources such as financial reports, surveys, press releases, news articles and social media. 

In this work, we fine-tune the approach proposed by~\citet*{Consoli2022a} employing the MoralStrength~\cite{Araque2020} and LibertyMFD~\cite{Araque2022154} lexicons to assess the morality expressed in news.
The extracted news sentiment and moral indicators are \emph{aspect-based} and \emph{fine-grained}, meaning that they are computed only for the specific topic of interest and expressed at a continuous value range. This allows adapting the sentiment and moral score as intensities regarding the topic of interest instead of the standard binary (``positive'' vs ``negative'') approach. 
We also extract references to geographic locations from the articles to perform a regional analysis.
We focus on gender inequalities in the labour market as seen through the Spanish press.

Social scientists have attested precarious labour market conditions for females \cite{Larranaga201283}, both at national (e.g., see \citet{Zolnik201376} in the United States, \citet{LewandowskaGwarda2018183} in Poland, or \citet{Tatli2021225} in Turkey), and regional \cite{Lester2001310} levels through digital media~\cite{urbinati2020young,rama2020facebook,berte2023monitoring}. 
Despite the problematic implications of this issue in Spain, the phenomenon remains mainly unstudied \citep{RePEc:mod:dembwp:0010, Campos-Soria20164143}. 

In this work, we aim to fill this gap by analysing how the Spanish news cover the female (un)employment topic in terms of sentiment and moral values, as well as how this sentiment evolves over time.
Our findings suggest a more negative trend relative to the extracted news sentiment and moral values for female unemployment when there is a labour market downturn. 
Moreover, this phenomenon is evident in the entire period of study (2000--2022) and at the national and regional levels. 

\section{Data Collection}

The data employed in this study originate from the Dow Jones News and Analytics platform (DNA, \url{https://www.dowjones.com/dna/}). 
In particular, we obtained the full-text news articles from the the main national Spanish outlets\footnote{\emph{El Mundo, ABC, Expansión, La Vanguardia, Cinco Días, El País - Nacional}} from January 2000 to December 2022, as well as for 31 regional ones\footnote{\emph{Actualidad Económica, Agencia EFE - Servicio Económico, Alimarket-Construcción, Alimarket-Envase, Alimarket-Non Food, Aseguranza, El Comercio, Córdoba, El Correo, El Diario de León, Diario Montañés, El Diario Vasco, Europa Press - Servicio Económico, Europa Press - Servicio en Catalán, Europa Press - Servicio en Gallego, Europa Press - Servicio en Valenciano, Europa Press - Servicio Nacional, Hoy, Ideal, El Mundo - Andalucía, El Mundo - Catalunya, El Mundo - País Vasco, El Mundo - Valencia, El Periódico de Aragón, El Periódico Extremadura, El Periódico Mediterráneo, Las Provincias, La Rioja, Sur, La Verdad, La Voz de Galicia}} from January 2000 to September 2019, resulting in a total of approximately 15 million articles. 
The choice of the outlets has been made to have a balanced representation of newspapers with respect to their political orientation, as well as to maximize the temporal and regional coverage. 
The articles cover a wide range of political and social issues, but exclude sports news since it is out of the scope of this study. 

For each article we obtain the publication date, the title and the body of the article given in Spanish. Given that our pipeline relies on the English language, we translate the articles from Spanish to English using \textit{eTranslation}, the machine translation service of the European Commission\footnote{\textit{eTranslation} service, available at \url{https://ec.europa.eu/info/resources-partners/machine-translation-public-administrations-etranslation_en}.}. 
For further details on the robustness of the \textit{eTranslation} approach in sentiment analysis, the reader is referred to the Appendix in \citet{barbaglia2022forecasting}, where the authors have carried out a Latent Dirichlet Allocation (LDA) analysis \citep{blei2003latent} of some original and translated texts in different languages, and found a quite satisfactory agreement score (around 80\% on average) between the different original and translated topics sets, suggesting an overall reliability and robustness of the translation process.

\section{Methodology}

In this study, we aim to extract the sentiment and moral values polarities from Spanish news to analyse female unemployment in the country. 
To do so, we employ the fine-grained aspect-based sentiment (FiGAS) approach \cite{Consoli2022a, Consoli202187, Consoli202052}\footnote{The algorithm is deployed on Python (version 3.7.6) and is accessible online from the original project \cite{Consoli2022a} at the GitHub repository: \url{https://github.com/sergioconsoli/SentiBigNomics}.}.
The FiGAS approach delivers indicators targeted to the topic of interest and accurately assesses the sentiment in a fine-grained and aspect-based manner. 
Here, we extend this method\footnote{we will reference the repository upon acceptance
} to include moral values polarities by employing the MoralStrength~\cite{Araque2020} and LibertyMFD~\cite{Araque2022154} dictionaries.

\textbf{Terms of Interest (ToI) Identification.} Given that we are interested in tracking news about the labour market and employment, we start from a set of related keywords, for instance, \emph{unemployment, labour market, employment growth, job opportunities, job market, salary}.
We further extend this list of terms using semantic synonyms, computed using the Sense2Vec python library\footnote{Sense2Vec library: \url{https://pypi.org/project/sense2vec/}.}, an extension to 
the popular Word2Vec model \citep{Mikolov2013} that learns the semantic similarities across all word vectors using a corpus of Reddit comments\footnote{Reddit: \url{https://www.reddit.com/}.} from the years 2015 to 2019. 
Given a keyword as input, Sense2Vec provides as output a set of synonyms with a semantic similarity score between zero and one. We set a relatively high threshold, selecting only synonyms relatively close in meaning to our input keywords, with a semantic similarity score of 0.7 or higher \citep{Colagrossi2022323}. 
Following this approach, we derived 362 keywords, including concepts like \emph{job creation, employment crisis, wage growth, unemployment benefits, payroll, lay-off, manpower}.\footnote{Full list available from the authors upon request.}

\textbf{Extraction of patterns.}
FiGAS~\cite{Consoli2022a,Consoli202187,Consoli202052} relies on the features provided by the \textit{spaCy} Python library\footnote{spaCy: Industrial-Strength Natural Language Processing in Python. Available at: \url{https://spacy.io/}}, namely, word vectors, context-specific token vectors, part-of-speech (POS) tags, dependency parse, and recognition of named entities. 
FiGAS loops over the part-of-speech tags, stopping when one of the terms of interest is found in a sentence associated with a geographic location of interest. 
Then, it navigates over the neighbouring tokens following a rule-based approach which leverages the syntactic dependency parsing. In this way, structured patterns related to the terms of interest in the desired location are extracted from the news article.

\textbf{Sentiment and Moral Lexicons.} We adopt a quantitative and a qualitative approach to perform sentiment analysis. In particular, we define the polarity (in the $[-1,+1]$ range) by employing a series of state-of-the-art lexicons such as SenticNet \cite{cambria2010senticnet}, SentiWordNet \cite{baccianella2010sentiwordnet}, or SentiBigNomics \cite{Consoli2022a}. 
Regarding morality, we employ the Moral Foundations Theory (MFT), which expresses the psychological basis of morality in terms of innate intuitions, defining the following five foundations: \textit{care/harm}, \textit{fairness/cheating}, \textit{loyalty/betrayal}, \textit{authority/subversion}, and \textit{purity/degradation} \cite[e.g.,][]{Haidt200455, Haidt200798}. 
MFT is broadly adopted in the computational social science field since it defines a clear taxonomy of values with direct application to a wide range of topics.
MFT explains worldviews around a variety of social issues, such as vaccine hesitancy~\cite{kalimeri2019human,gaston2022moral} and the emergence of symbolism in resistance movements \cite{mejova2023authority} and has also proven useful in understanding narratives and attitudes towards unemployment~\cite{bonanomi2017understanding,urbinati2020young} and general press biases~\cite{kuypers2020president}.

In this work, we employ the \emph{MoralStrength} lexicon~\cite{Araque2020}, which provides for each lemma a numeric assessment of Moral Valence, indicating the strength with which the lemma expresses the specific value. 
We also incorporate the most recent LibertyMFD lexicon~\cite{Araque2022154} to assess the moral foundation of \textit{liberty}.
The transparency of a lexicon-based approach represents a significant advantage in computational social science applications providing policymakers and stakeholders with a direct interpretation of the drivers behind the models' predictions.
 
\textbf{Polarity scoring and propagation.}
When a pattern is detected, FiGAS assigns a polarity score to each of the tokens in the pattern.
The polarity scores are assigned according to the lexicons of sentiment and moral values~\footnote{Note that the valence scores, originally reported in a 1 to 9 scale, are rescaled in the $[-1,+1]$ continuous range for consistency of the analysis.} respectively, resulting in a seven-dimensional feature space. 
Using a polarity propagation mechanism \cite{Consoli2022a}, the scores are propagated to the root term of interest (ToI) detected in the pattern, producing an overall polarity score for the ToI per geographical location.

For more details and features of the algorithm, its overall pipeline, the syntactic and semantic rules adopted to build the patterns, and the polarity scoring and propagation mechanisms, the reader is referred to \cite{Consoli2022a, Consoli202187, Consoli202052} for a detailed presentation.

\textbf{Geographic Analysis.} To provide a geographic breakdown of the sentiment and moral values to enrich the insights around the female (un)employment topic, we add another component to the analysis. From each article, we extract the reference to a specific \textit{geographic location} in Spain.
In case no location is identified ($42\%$ of cases), we assume that it refers to the main location mentioned in the entire article text. 
The analysis is performed at Nomenclature of Territorial Units for Statistics (NUTS), a geocode standard for referencing the administrative divisions of countries for statistical purposes.
In particular, here we employ the NUTS3 regional level\footnote{
See \url{https://ec.europa.eu/eurostat/web/nuts/nuts-maps}} which corresponds to provinces, 
along with the related synonyms, although the final results are reported at NUTS2 level (i.e. region names\footnote{\emph{Andalusia, Aragon, Balearic islands, Basque country, Canary islands, Cantabria, Castilla la Mancha, Castile Leon, Catalonia, Ceuta
Extremadura, Galicia, la Rioja, Madrid, Melilla, Navarre, principality of Asturias, region of Murcia, Valencian community}}) to simplify the exposition. 

\begin{figure}[!ht]
 \centering
 \includegraphics[width=0.48\textwidth]{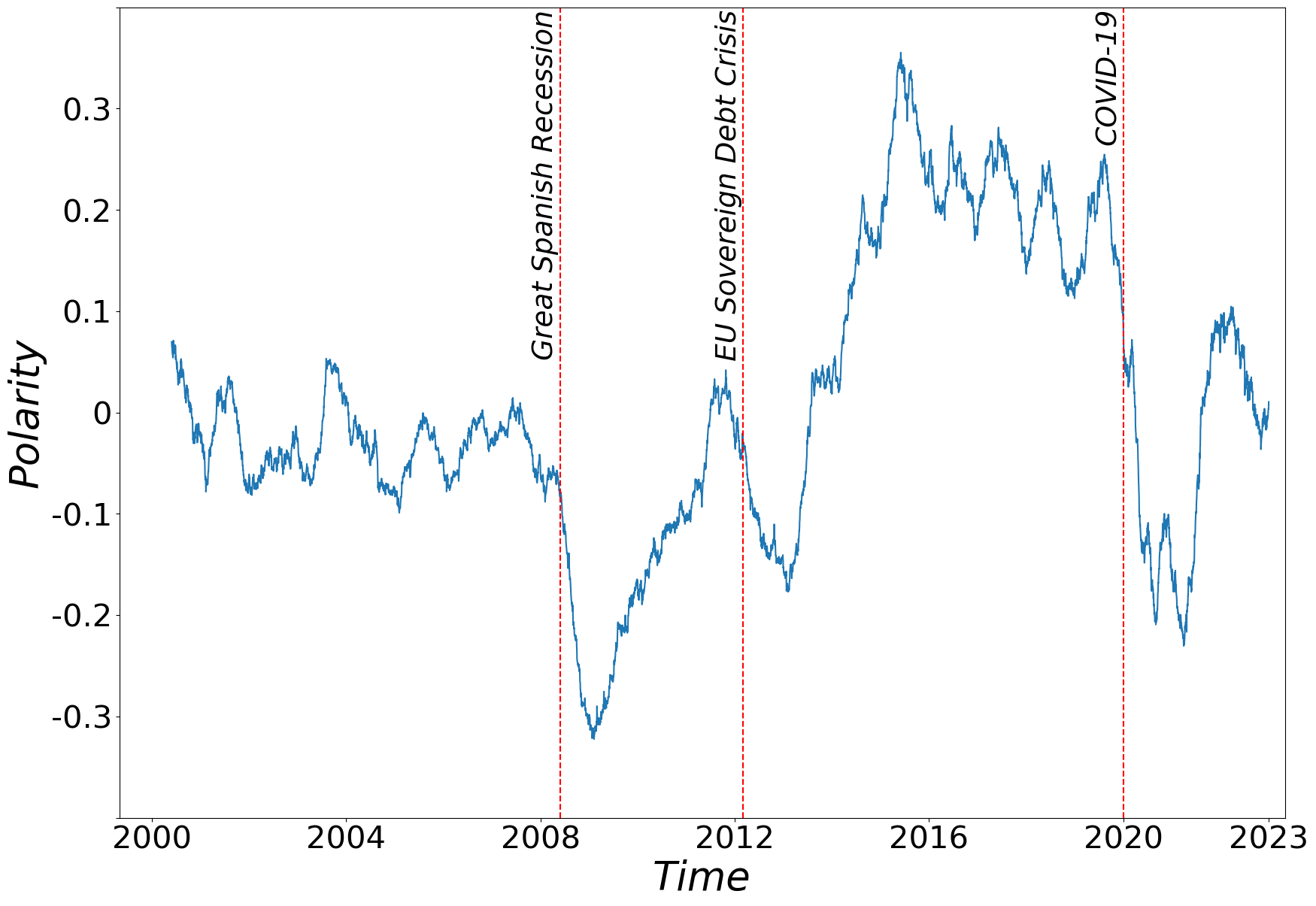} 
 \caption{Articles' sentiment for the daily news about the \emph{(un)employment} and \emph{labour market} topics in \emph{Spain} 
 using a 90 days smoothing window. The average number of observations is normalised by the number of articles released over the same period. Red dotted lines indicate the beginnings of important periods of economic and social distress (\emph{Great Spanish Recession}, \emph{EU Sovereign Debt Crisis}, and \emph{COVID-19 pandemic}).}
 \label{fig:sentiment_spain}
\end{figure}

\section{Results \& Discussion}

Initially, all Spanish news articles are automatically translated into English employing eTranslation. 
Employing the FiGAS approach we extract the segments of interest, filter 
the mentioned locations at the Spanish NUTS3 level for the labour market and employment topics, and extract the sentiment and moral values polarity scores. 
Overall, we obtained 678,276 patterns emerging from 369,216 unique articles out of the $\approx 15$ million unique Spanish news. 

Polarity scores are grouped per sentiment and moral dimension by averaging the calculated polarity scores daily.
For consistency, the average polarities are normalised 
by the number of articles and outlets available in each sample period. Figure \ref{fig:sentiment_spain} depicts the articles' sentiment for the news. 
We notice a significant downtrend of the articles' sentiment at the end of 2008, coinciding with the bankruptcy of Lehman Brothers that triggered the global financial crisis\footnote{\url{https://www.federalreservehistory.org/essays/great-recession-and-its-aftermath}}. This event corresponds to the most prominent retention in the history of the Spanish economy, also known as the Great Spanish Recession \cite{RePEc:fda:fdaeee:eee2016-10}, where the Spanish IBEX-35 stock exchanges index fell by more than 9\% in a single day, at the end of October 2008\footnote{\url{https://www.rabobank.com/knowledge/d011323133-the-crisis-and-recovery-of-spains-labour-market}}. 
A similar yet smaller decrease in overall sentiment can be seen after 2012, which coincides
with the EU bond crisis \citep{Lane2012}, which brought the Spanish economy again to its knees \citep{eubondSpainRep}.
We can also detect a significant decrease
in 2020, which marks the beginning of the COVID-19 pandemic.
The articles' negative tone during the first phase of the pandemic could depend on a generalised concern towards employment in the overall Spanish economy but also towards the specific adverse impact that the pandemic had on the entire labour market. 

The ``speed of recovery" made the sentiment rise again after the first wave of the pandemic, probably due to the introduction of the vaccines and support measures for companies and workers, which helped generate collective confidence in a ``post-COVID" labour market future. 
For instance, for the following example snippet: ``{\it Spanish workers call for purchasing power to be maintained in 1994 and 1995 and employers propose to freeze salaries this year.}", the ToI ``{\it salary}" ignites the pattern ``{\it freeze salaries in...}" which propagates a -0.28 sentiment polarity score to the ToI. Note that the algorithm would recognise in the sentence the adjectival modifier {\it Spanish}; in this case, the entity recognition of the algorithm detects a nationality that is used to derive the location ``Spain" to assign to the pattern. As another example of sentiment detection, the algorithm recognises the ToI ``{\it job insecurity}" and location ``{\it Catalunia}" from the sentence. 
Another example snippet reads: ``{\it The report indicates that the main differences in the female employment relationship between females in the labour market in Catalonia are greater job insecurity and difficult access to the first job.}", igniting the pattern: ``{\it greater job insecurity}" which results in -0.77 sentiment polarity score. In this example, the algorithm also detects the explicit reference to the female gender for the labour market.

\begin{table}[!t]
\begin{center}
\setlength{\tabcolsep}{3pt}
\begin{footnotesize}
\centering
 \begin{tabular}{lccc}
 \toprule
 \textbf{} & \textbf{Female} & & \textbf{Non-Female} \\
 \midrule
count &	16,899	& & 632,662 \\
mean &	0.062	& & 0.069 \\
std	 & 0.479 & & 0.420 \\
10\% &	\bf{-0.662} & & -0.493 \\
25\% &	\bf{-0.321} & & -0.220 \\
50\% &	0.108 & & 0.116 \\
75\% &	0.383 & & 0.334 \\
90\% &	0.741 & & 0.601 \\
\bottomrule
\end{tabular}
\caption{Overall statistics of the sentiment polarity between the \emph{(un)employment} and \emph{labour market} topics for ``Female'' and ``Non-Female''. A Mann-Whitney U test on the two distributions show that they are not statistically different. A Mann-Whitney U test on the 10\% and 25\% quantiles showed a statistical significant difference in those segments.}
\label{tab:stats-sentiment}
\end{footnotesize}
\end{center}
\end{table}

Turning to the female employment and labour market, we filter segments with explicit female references and compare them with the remaining ones, i.e. referred to males or those of unspecified gender.
Table \ref{tab:stats-sentiment} summarises those statistics for the two groups, referred to as, respectively, ``Female'' and ``Non-Female''. Observing the mean and standard deviation, we do not detect a remarkable difference between the two time series. This is also confirmed if we perform a Mann-Whitney U test \citep{Neuhauser20111} on the mean of the two distributions, testing the null hypothesis that their samples are equally distributed. We obtain a \textit{U}-statistic = 34,733 and a \textit{p}-value = 0.07 which, being it larger than 0.05, indicates weak evidence against the null hypothesis. As a result, the null hypothesis is not rejected at a 95\% significance level, and we can state that the mean of the two distributions is not statistically different.

But if we look at the first decile of the distributions in Table \ref{tab:stats-sentiment}, the sentiment of ``Female'' looks quite lower than that of ``Non-Female'' (values in bold in the table); the same holds for the 25\% segment. Performing the Mann-Whitney U test in those cases, we find that ``Female'' have statistically lower sentiment scores than the ``Non-Female'' (\textit{U}-statistic = 22,684 and a \textit{p}-value = $9.62 \cdot 10^{-17}$ for the 10\% quantile, and a \textit{U}-statistic = 25,392 and a \textit{p}-value = $6.15 \cdot 10^{-12}$ for the 25\% quantile).
This indicates a strong difference in the lower tails of the distributions, indicating a difference in outliers in reporting between the genders.
This shows that news is reporting a more negative sentiment when referring to female 
labour market during periods of high uncertainty and distress, meaning that females appear to be affected more in those cases.

\begin{table}[!t]
\begin{center}
\setlength{\tabcolsep}{3pt}
\begin{footnotesize}
 \centering
 \begin{tabular}{lcccccc}
 \toprule
\textbf{year}	&	\textbf{mean}	&	\textbf{10\%}	&	\textbf{25\%}	&	\textbf{50\%}	&	\textbf{75\%}	&	\textbf{90\%}	\\
 \midrule
2000	&	1.78	&	\bf{-38.17}	&	\bf{-8.81}	&	5.44	&	15.07	&	34.04	\\
2001	&	5.57	&	\bf{-43.12}	&	\bf{-14.82}	&	3.85	&	25.97	&	57.48	\\
2002	&	-0.34	&	\bf{-52.05}	&	\bf{-16.17}	&	1.54	&	21.28	&	46.23	\\
2003	&	3.43	&	\bf{-31.30}	&	\bf{-16.10}	&	2.03	&	25.60	&	43.15	\\
2004	&	1.30	&	\bf{-51.70}	&	\bf{-16.25}	&	3.11	&	22.06	&	50.88	\\
2005	&	-1.52	&	\bf{-33.38}	&	\bf{-19.45}	&	-4.04	&	18.71	&	31.47	\\
2006	&	-1.86	&	\bf{-44.17}	&	\bf{-15.72}	&	-3.40	&	14.26	&	34.94	\\
2007	&	2.64	&	\bf{-30.85}	&	\bf{-11.12}	&	1.83	&	19.62	&	36.92	\\
2008	&	-2.22	&	\bf{-32.89}	&	\bf{-16.07}	&	-1.20	&	12.70	&	32.29	\\
2009	&	-1.69	&	\bf{-50.09}	&	\bf{-18.72}	&	-1.25	&	16.96	&	42.74	\\
2010	&	0.69	&	\bf{-36.22}	&	\bf{-15.17}	&	2.03	&	17.06	&	40.01	\\
2011	&	-0.20	&	\bf{-59.92}	&	\bf{-23.58}	&	0.97	&	22.83	&	46.58	\\
2012	&	2.05	&	\bf{-39.70}	&	\bf{-15.16}	&	2.40	&	20.07	&	37.91	\\
2013	&	-3.36	&	\bf{-40.57}	&	\bf{-22.83}	&	-2.89	&	20.20	&	36.65	\\
2014	&	1.90	&	\bf{-44.06}	&	\bf{-21.33}	&	1.09	&	24.03	&	50.58	\\
2015	&	-3.08	&	\bf{-42.37}	&	\bf{-25.00}	&	-3.12	&	17.73	&	32.23	\\
2016	&	-8.70	&	\bf{-46.37}	&	\bf{-27.53}	&	-6.46	&	7.66	&	32.83	\\
2017	&	-3.19	&	\bf{-42.84}	&	\bf{-20.64}	&	-2.34	&	15.72	&	34.00	\\
2018	&	-5.99	&	\bf{-46.14}	&	\bf{-23.19}	&	-3.87	&	11.66	&	24.74	\\
2019	&	-1.55	&	\bf{-33.96}	&	\bf{-16.06}	&	-2.97	&	15.90	&	31.41	\\
2020	&	-9.17	&	\bf{-82.59}	&	\bf{-36.81}	&	-4.96	&	20.22	&	37.76	\\
2021	&	-13.96	&	\bf{-77.43}	&	\bf{-32.58}	&	-13.79	&	12.99	&	36.52	\\
2022	&	-3.76	&	\bf{-52.74}	&	\bf{-25.37}	&	0.05	&	16.64	&	39.10	\\
\bottomrule
 \end{tabular}
 \caption{Percentage difference in sentiment polarity per \emph{year} between the \emph{(un)employment} and \emph{labour market} topics for ``Female'' and ``Non-Female''.}
 \label{tab:perce-difference-year-sentiment}
 \end{footnotesize}
 \end{center}
\end{table}

\begin{figure}[!t]
 \centering
 \includegraphics[width=0.45\textwidth]{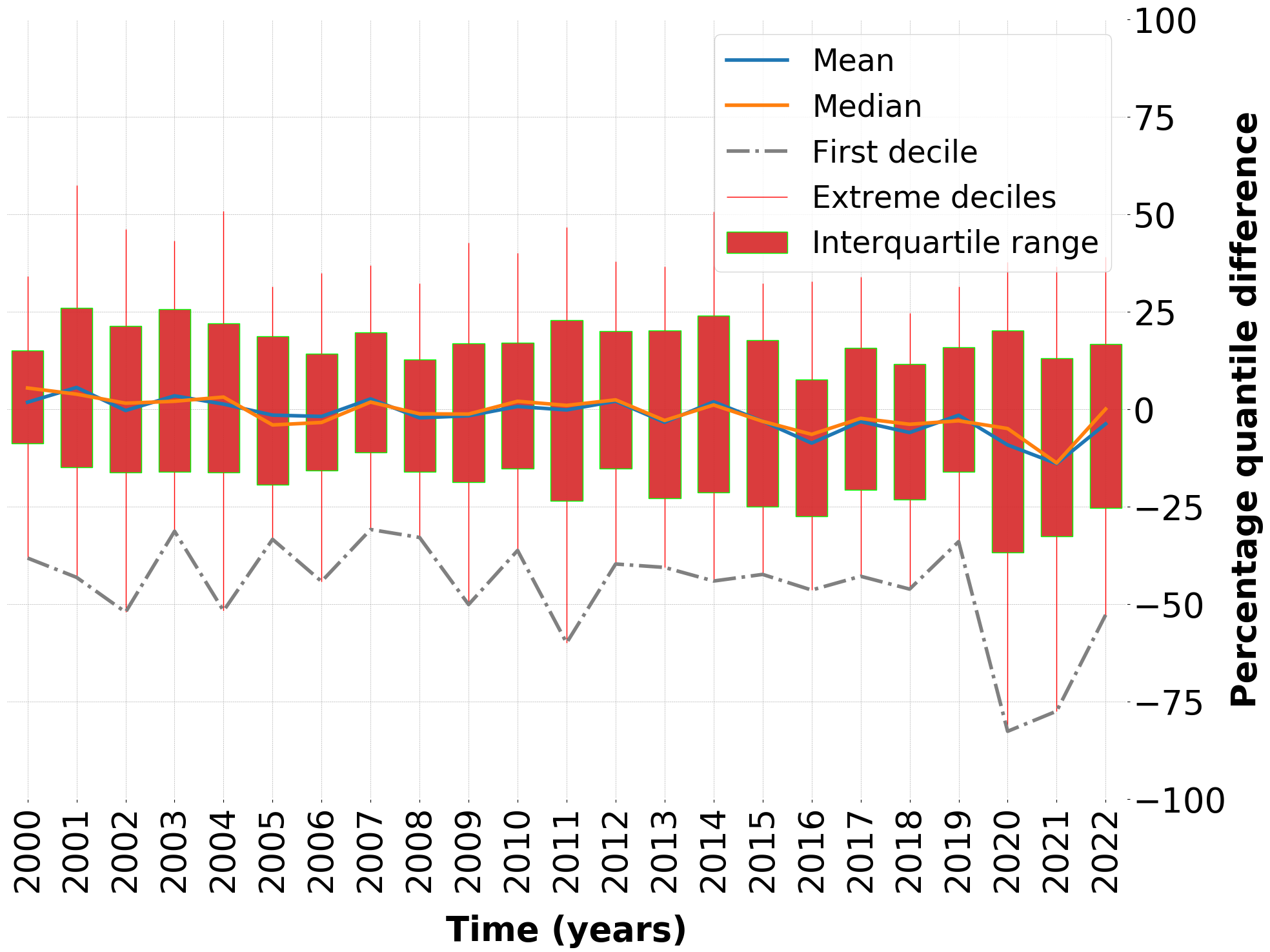} 
 \caption{Sentiment polarity distributions per year (see Table \ref{tab:perce-difference-year-sentiment}) The dashed line corresponds to the first decile.}
 \label{fig:df_diff_percentage_By_YEAR-candlestick_chart2-sentiment}
\end{figure}

Focusing on the time evolution of the phenomenon, Table \ref{tab:perce-difference-year-sentiment} 
reports the yearly percentage change in sentiment polarity for the topics of interest between ``Female'' and ``Non-Female'', highlighting the low quantiles (10\% and 25\%) of the percentage change. 
Figure \ref{fig:df_diff_percentage_By_YEAR-candlestick_chart2-sentiment} visually depicts the same information clearly showing the impact of the Great Spanish Recession, with a drop of -50.09\% for 2009 at the 10\% quantile, and the impact of the COVID-19 pandemic, with a drop of -82.59\% for 2022 and -77.43\% for 2021 at the 10\% quantile, respectively.

\begin{figure}[!h]
 \centering
 \includegraphics[width=0.47\textwidth]{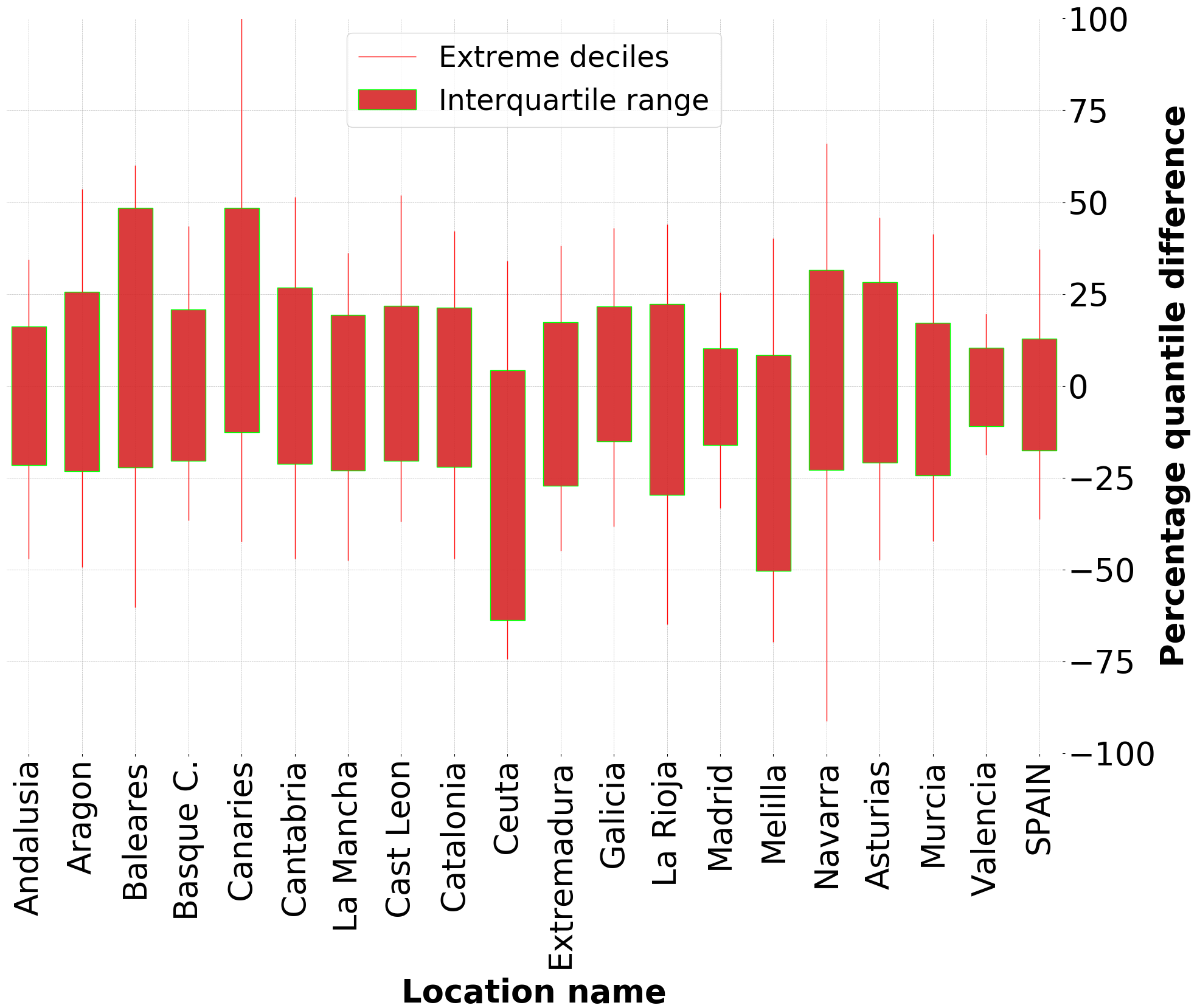} 
 \caption{Regional (NUTS2) percentage quantile differences in sentiment polarity (see Table \ref{tab:perce-difference-nuts2-sentiment})}
 \label{fig:df_diff_percentage_By_NUTS2-candlestick_chart2-sentiment}
\end{figure}

\begin{figure}[!h]
 \centering
 \includegraphics[width=0.47\textwidth]{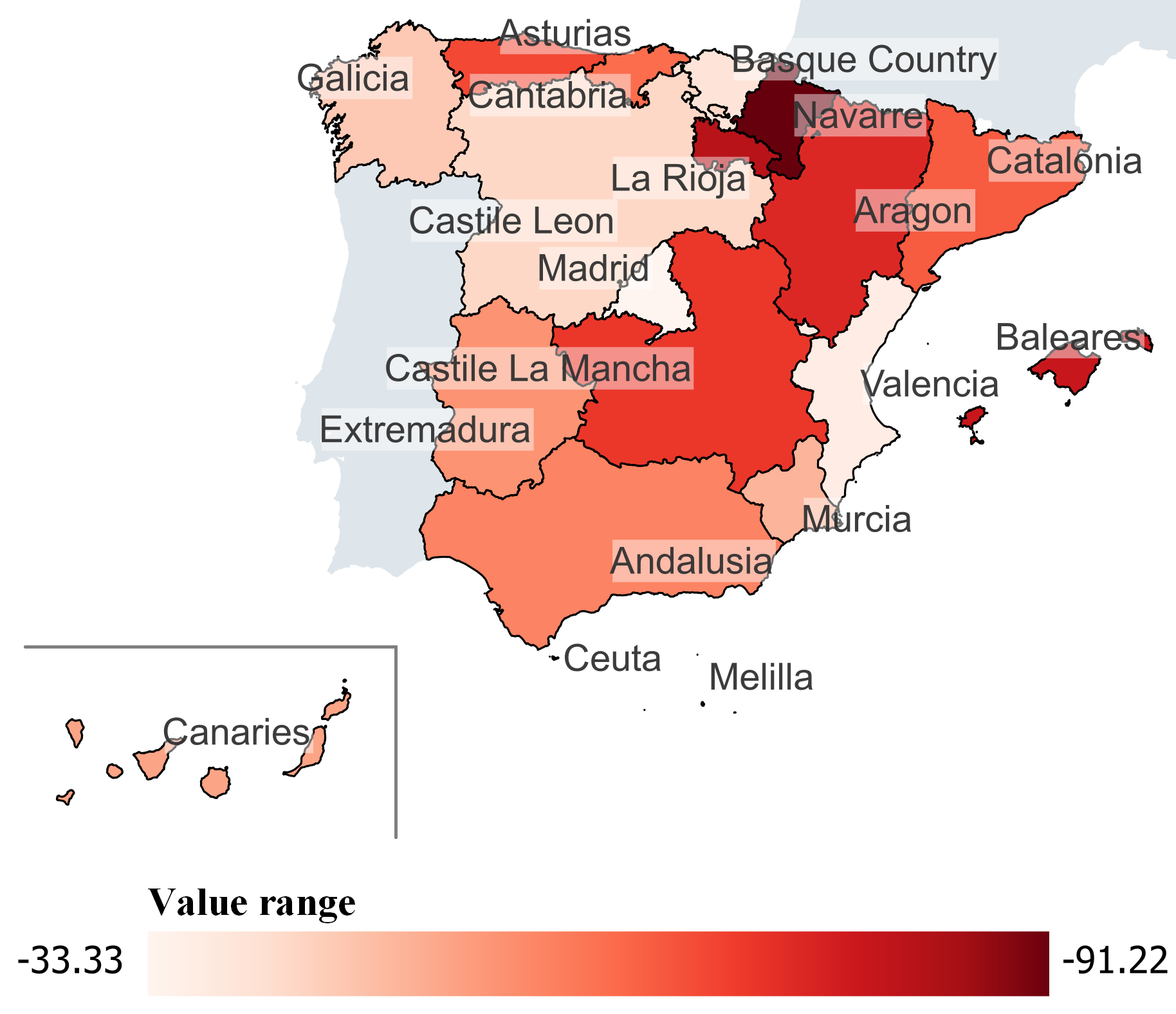} 
 \caption{Geographic distribution of the percentage difference in sentiment polarity (10\% quantile) for each Spanish region (\emph{NUTS2}) between the \emph{(un)employment} and \emph{labour market} topics for ``Female'' and ``Non-Female''.
A brighter red shade indicates an higher percentage difference and hence a more negative sentiment on the way the ``Female'' population is discussed about in the labour market related press.}
 \label{fig:sentiment_nuts2}
\end{figure}

Turning to the geographic distribution of the inequalities, Table \ref{tab:perce-difference-nuts2-sentiment} reports the changes in sentiment polarity per region (NUTS2 level), with the highlighted values in the Table \ref{tab:perce-difference-nuts2-sentiment} to indicate the 10\% and 25\% quantiles. 
Figure \ref{fig:df_diff_percentage_By_NUTS2-candlestick_chart2-sentiment} depicts the distributions of sentiment per region, providing a visual rappresentation of the regions where the the difference in sentiment polarity, and hence, the way articles talk about ``Female'' in the labour market is more negative.
Figure \ref{fig:sentiment_nuts2} shows the geographical distribution of the phenomenon focusing on the first decile. 
In general Spanish regions report worse values at low quantiles with respect to the national average (referred to as SPAIN in the table and in the candlestick chart). The most noticeable drops in the first decile appear in smaller regions, like the Balearic Islands, Melilla, Navarre, Ceuta, and La Rioja, perhaps caused by a major focus on local labour markets rather than on aggregate statistics, or because in local contexts females might undergo isolation or discrimination cases more often. 
This confirms our previous findings on the hardships of females reported in the Spanish labour market.


\begin{table}[!h]
\begin{center}
\setlength{\tabcolsep}{3pt}
\begin{footnotesize}
 \centering
 \begin{tabular}{lcccccc}
 \toprule
\textbf{NUTS2}	&	\textbf{mean}	&	\textbf{10\%}	&	\textbf{25\%}	&	\textbf{50\%}	&	\textbf{75\%}	&	\textbf{90\%}	\\
 \midrule
Andalusia	&	-6.05	&	\bf{-47.04}	&	\bf{-21.61}	&	-6.76	&	16.16	&	34.43	\\
Aragon	 &	3.24	&	\bf{-49.44}	&	\bf{-23.20}	&	5.95	&	25.66	&	53.53	\\
Baleares	&	4.56	&	\bf{-60.25}	&	\bf{-22.21}	&	11.25	&	48.36	&	59.96	\\
Basque C.	 &	0.98	&	\bf{-36.59}	&	\bf{-20.41}	&	-0.21	&	20.80	&	43.47	\\
Canaries	&	19.07	&	\bf{-42.43}	&	\bf{-12.65}	&	3.26	&	48.51	&	108.35	\\
Cantabria	&	5.23	&	\bf{-47.05}	&	\bf{-21.32}	&	8.72	&	26.77	&	51.46	\\
La Mancha	&	-2.41	&	\bf{-47.52}	&	\bf{-23.02}	&	-0.89	&	19.34	&	36.15	\\
Cast Leon	&	3.73	&	\bf{-36.99}	&	\bf{-20.50}	&	5.46	&	21.71	&	51.82	\\
Catalonia	&	-2.58	&	\bf{-47.14}	&	\bf{-22.11}	&	-1.90	&	21.22	&	42.09	\\
Ceuta	 &	-25.37	&	\bf{-74.41}	&	\bf{-63.80}	&	-21.41	&	4.17	&	34.00	\\
Extremadura	&	-5.10	&	\bf{-44.99}	&	\bf{-27.15}	&	-3.48	&	17.36	&	38.20	\\
Galicia	 &	2.14	&	\bf{-38.33}	&	\bf{-15.20}	&	2.09	&	21.54	&	42.93	\\
La Rioja	&	-4.77	&	\bf{-64.95}	&	\bf{-29.61}	&	-2.11	&	22.33	&	44.00	\\
Madrid	 &	-4.42	&	\bf{-33.33}	&	\bf{-16.19}	&	-2.18	&	10.25	&	25.41	\\
Melilla	 &	-13.67	&	\bf{-69.75}	&	\bf{-50.29}	&	-9.33	&	8.45	&	40.17	\\
Navarra	 &	1.96	&	\bf{-91.22}	&	\bf{-22.95}	&	28.84	&	31.52	&	66.02	\\
Asturias	&	3.80	&	\bf{-47.32}	&	\bf{-20.96}	&	9.87	&	28.23	&	45.74	\\
Murcia	 &	-2.14	&	\bf{-42.34}	&	\bf{-24.45}	&	-1.14	&	17.21	&	41.24	\\
Valencia	&	-0.96	&	\bf{-36.32}	&	\bf{-17.64}	&	-3.06	&	12.86	&	37.14	\\
 \midrule
SPAIN	 &	0.28	&	\bf{-18.74}	&	\bf{-10.99}	&	0.65	&	10.31	&	19.58	\\
\bottomrule
 \end{tabular}
 \caption{Percentage difference in sentiment polarity for each Spanish region (\emph{NUTS2}) and for the entire country (\emph{SPAIN}) between the \emph{(un)employment} and \emph{labour market} topics for ``Female'' and ``Non-Female''.}
 \label{tab:perce-difference-nuts2-sentiment}
 \end{footnotesize}
 \end{center}
\end{table}

\begin{figure*}[!t]
\centering
\caption{Percentage difference (10\% quantile) in polarity per year for the \textit{liberty} and \textit{fairness} moral values between ``Female'' and ``Non-Female'' for the \emph{(un)employment} and \emph{labour market} topics.}
\label{fig:diff_quantiles}
 \subfloat[Liberty]{
 \includegraphics[width=0.49\linewidth]{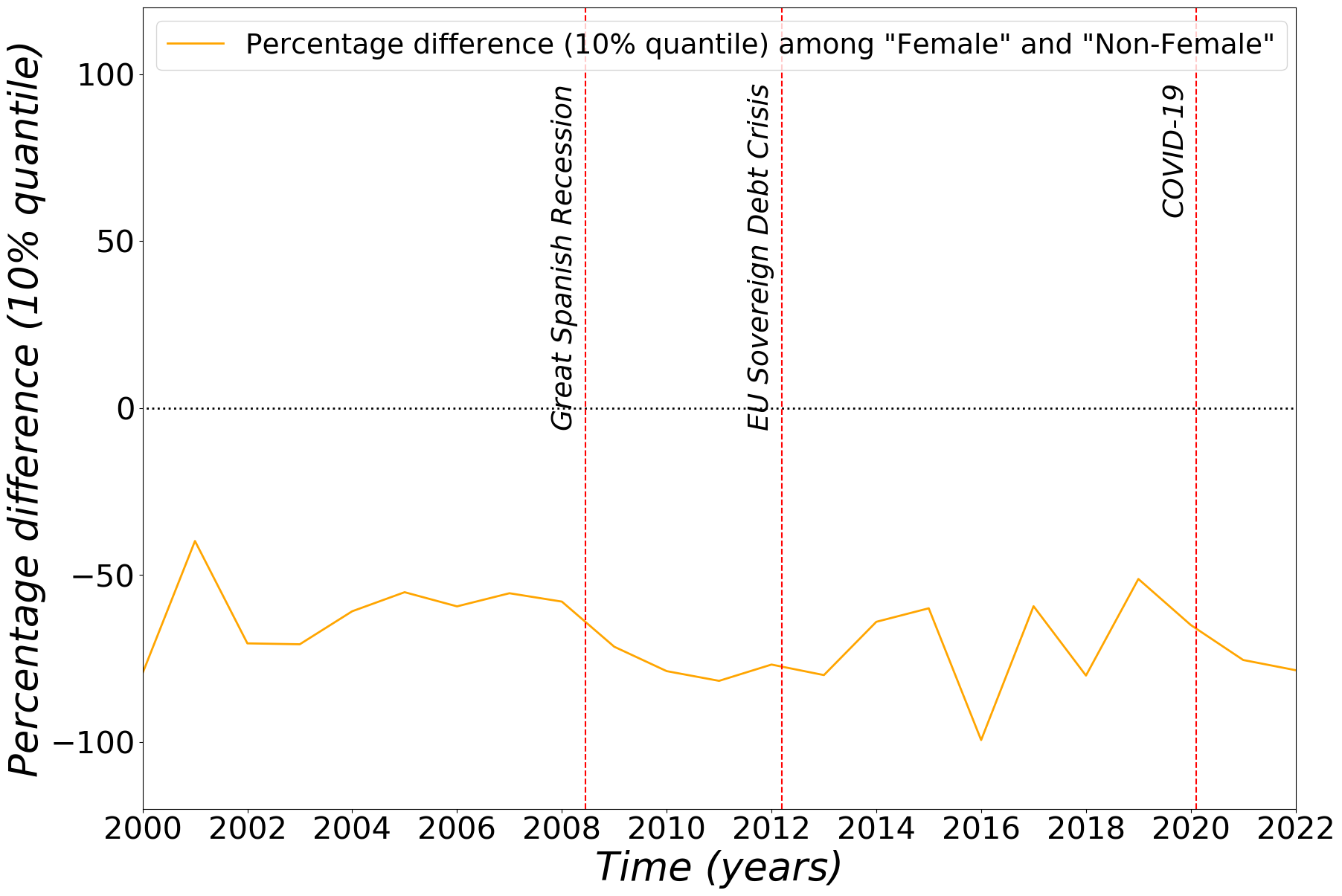}\label{fig:Liberty}}
 \subfloat[Fairness]{
 \includegraphics[width=0.49\linewidth]{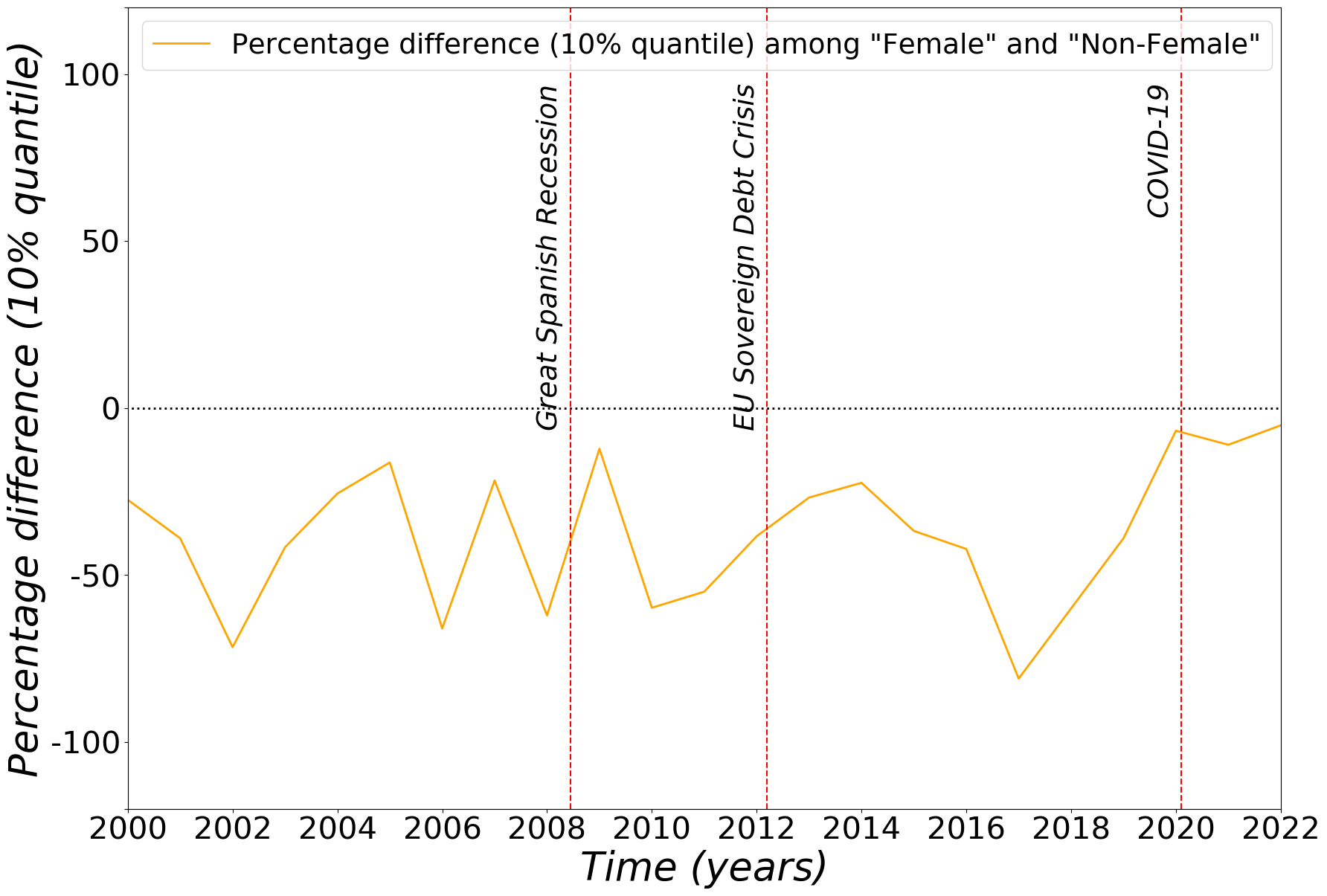}\label{fig:Fairness}}
\end{figure*}

\begin{figure}[!t]
 \centering
 \includegraphics[width=0.47\textwidth]{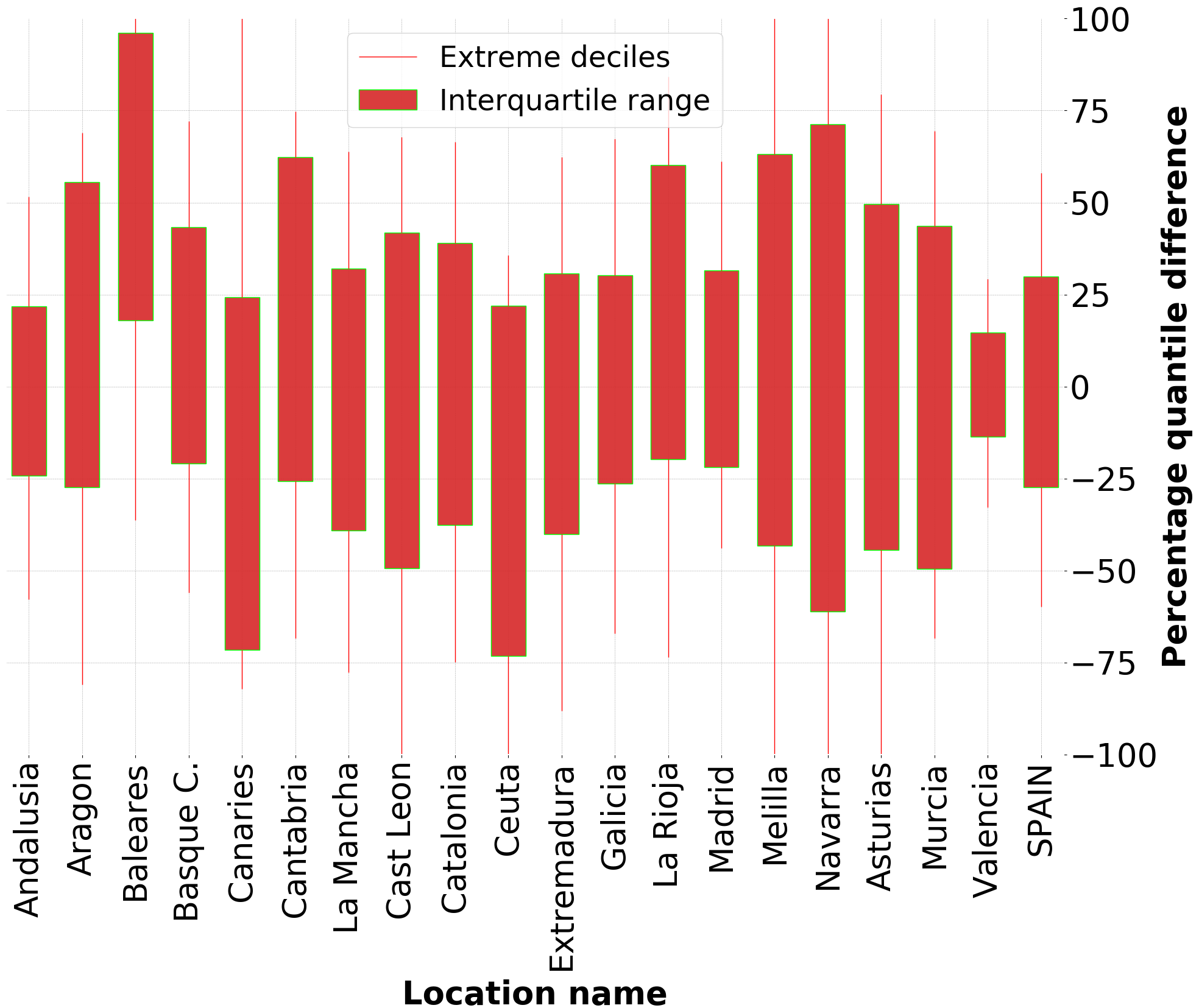} 
 \caption{Percentage difference distributions per Spanish regions (NUTS2) (see Table \ref{tab:perce-difference-nuts2-liberty}).}
 \label{fig:df_diff_percentage_By_NUTS2-candlestick_chart2-liberty}
\end{figure}

Turning to the moral values, we observe a similar behaviour; the low quantiles (10\% and 25\%) of the percentage polarity differences between the ``Female'' and ``Non-Female'' distributions remarkable drops. The phenomenon is more prominent for the moral values of \textit{liberty} and \textit{fairness}. 
In the case of liberty, an excerpt would read: ``{\it Prejudices are preventing females from entering posts traditionally reserved for males.}'', the algorithm recognises the ToI ``{\it post}'' and ignites the pattern: ``{\it preventing females from entering posts traditionally reserved...}''; the final result, indicates a -0.53 polarity score for liberty. In this example, the location is not specified in the sentence; therefore, we assume that it refers to the central location mentioned in the entire article text (being ``{\it Madrid}'' in this particular example). 
To better understand fairness, we provide an excerpt of the text ``{\it In the Basque Country, young people and females workers also seem to have finally acquired the right of being able to enter the labour market.}'', the ToI ``{\it females workers}'' triggers the pattern ``{\it young people and females workers also seem to have finally acquired the right...}'' which would produce a 0.75 polarity score for the value of fairness associated to the ToI, also associated to the detected ``{\it Basque Country}'' location.
Figure \ref{fig:diff_quantiles} reports the first decile of the percentage change in polarity per year for the \textit{liberty} and \textit{fairness} moral values between the (un)employment and labour market topics for ``Female'' and ``Non-Female''\footnote{Here reported only \textit{fairness} and \textit{liberty} for spacing purposes.}. All the curves are below the zeros, meaning that female (un)employment scored a lower moral value polarity at peaks of the polarity distributions.
In Table \ref{tab:perce-difference-nuts2-liberty}, we report the percentage change in polarity for the \emph{Liberty} moral value per Spanish region (NUTS2) as well as for the entire country (SPAIN) of articles discussing about the labour market topics regarding the ``Female'' and ``Non-Female'' populations.
Figure \ref{fig:df_diff_percentage_By_NUTS2-candlestick_chart2-liberty} depicts the respective distributions.  
In line with the percentage change in sentiment polarity reported in Table \ref{tab:perce-difference-nuts2-sentiment}, several Spanish regions account lower scores for \emph{liberty} with respect to the national scores (see ``SPAIN'' in the Table \ref{tab:perce-difference-nuts2-liberty}), 
showing cases where females are more oppressed and probably dealing with unfairness. 

\begin{table}[!t]
\begin{center}
\setlength{\tabcolsep}{3pt}
\begin{footnotesize}
 \centering
 \begin{tabular}{lcccccc}
 \toprule
\textbf{NUTS2}	&	\textbf{mean}	&	\textbf{10\%}	&	\textbf{25\%}	&	\textbf{50\%}	&	\textbf{75\%}	&	\textbf{90\%}	\\
 \midrule
Andalusia	&	-2.94	&	\bf{-57.76}	&	\bf{-24.17}	&	-0.76	&	21.79	&	51.50	\\
Aragon 	 &	4.72	&	\bf{-80.93}	&	\bf{-27.43}	&	11.35	&	55.55	&	68.91	\\
Baleares	&	48.88	&	\bf{-36.29}	&	17.97	&	65.84	&	96.07	&	115.56	\\
Basque C.	&	6.76	&	\bf{-56.08}	&	\bf{-20.96}	&	13.05	&	43.26	&	72.13	\\
Canaries	&	-4.44	&	\bf{-82.15}	&	\bf{-71.57}	&	-7.35	&	24.21	&	109.82	\\
Cantabria	&	10.94	&	\bf{-68.48}	&	\bf{-25.78}	&	20.13	&	62.41	&	74.79	\\
La Mancha	&	-5.46	&	\bf{-77.76}	&	\bf{-39.20}	&	-4.96	&	31.97	&	63.82	\\
Cast Leon	&	-9.70	&	\bf{-103.42}	&	\bf{-49.45}	&	-4.55	&	41.73	&	67.73	\\
Catalonia	&	-1.86	&	\bf{-74.90}	&	\bf{-37.68}	&	3.20	&	39.06	&	66.53	\\
Ceuta 	 &	-31.81	&	\bf{-129.70}	&	\bf{-73.26}	&	-13.72	&	21.89	&	35.71	\\
Extremadura	&	-8.77	&	\bf{-88.08}	&	\bf{-40.19}	&	-5.92	&	30.72	&	62.36	\\
Galicia 	&	-0.16	&	\bf{-67.11}	&	\bf{-26.42}	&	4.08	&	30.20	&	67.23	\\
La Rioja	&	11.66	&	\bf{-73.54}	&	\bf{-19.82}	&	19.46	&	60.10	&	84.18	\\
Madrid 	&	4.40	&	\bf{-43.84}	&	\bf{-21.89}	&	7.15	&	31.50	&	61.13	\\
Melilla 	&	2.78	&	\bf{-121.61}	&	\bf{-43.23}	&	-10.71	&	63.22	&	124.50	\\
Navarre 	&	-13.76	&	\bf{-137.91}	&	\bf{-61.06}	&	-21.46	&	71.30	&	107.06	\\
Asturias	&	-3.23	&	\bf{-102.20}	&	\bf{-44.35}	&	0.56	&	49.57	&	79.45	\\
Murcia 	 &	-2.32	&	\bf{-68.49}	&	\bf{-49.60}	&	-2.61	&	43.66	&	69.39	\\
Valencia	&	-0.93	&	\bf{-59.86}	&	\bf{-27.41}	&	-0.06	&	29.88	&	58.08	\\
\midrule
SPAIN	&	-1.74	&	\bf{-32.81}	&	\bf{-13.63}	&	0.45	&	14.72	&	29.22	\\
\bottomrule
 \end{tabular}
 \caption{Percentage difference in polarity for the ``Liberty'' moral value extracted for each Spanish region (\emph{NUTS2}) and for the entire country (\emph{SPAIN}) between the \emph{(un)employment} and \emph{labour market} topics for ``Female'' and ``Non-Female''.}
 \label{tab:perce-difference-nuts2-liberty}
 \end{footnotesize}
 \end{center}
\end{table}

\begin{figure*}[!ht]
\centering
\caption{Zoom on the articles' polarity during the COVID-19 period for the extracted \emph{sentiment} and \emph{liberty} moral value for the \emph{(un)employment} and \emph{labour market} topics for ``Female'' (in red) and ``Non-Female'' (in blue), using a 90 days smoothing window.}
\label{fig:covid_zoom_sentiment_liberty}
 \subfloat[Sentiment during COVID-19]{
 \includegraphics[width=0.49\linewidth]{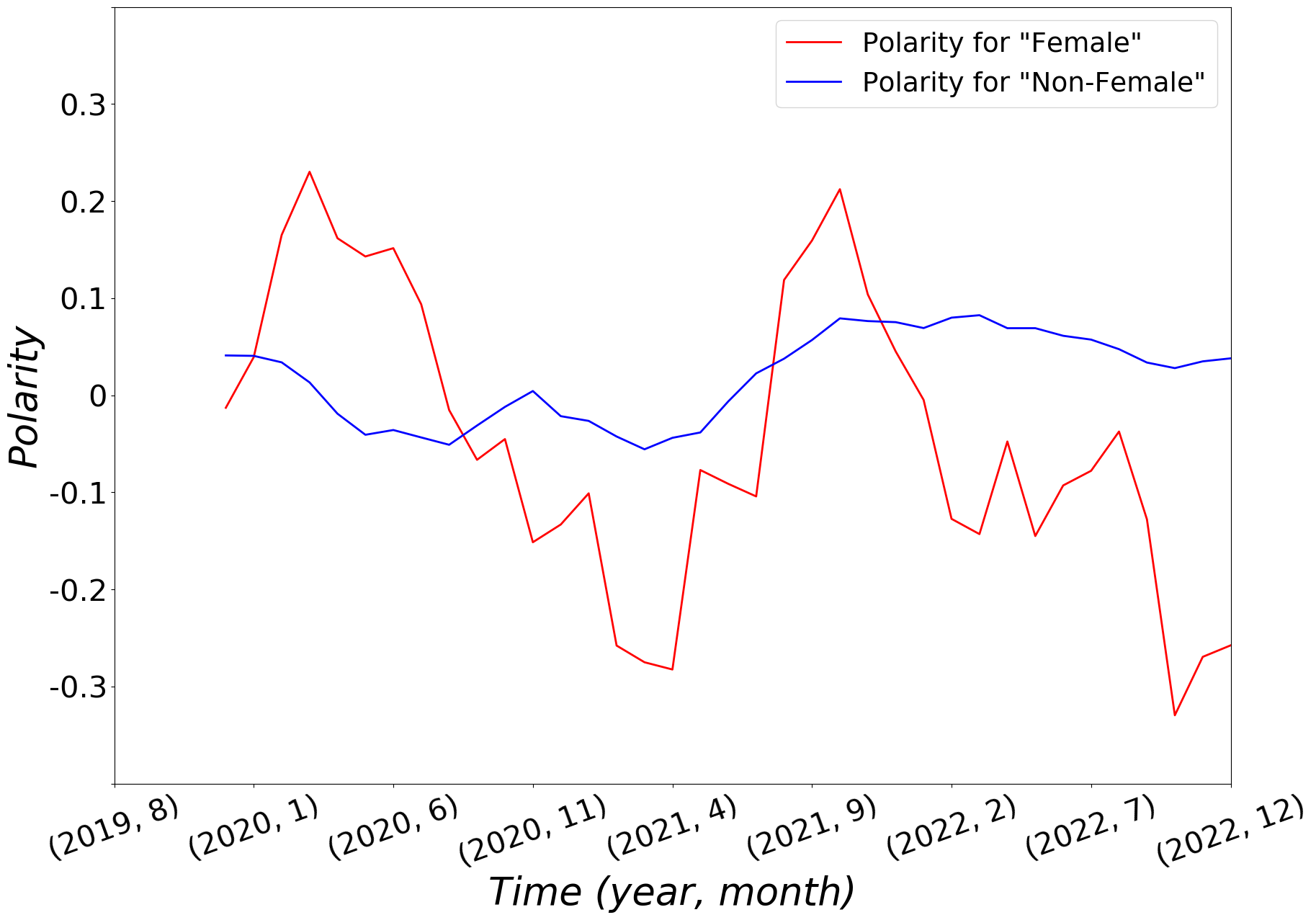}\label{fig:covid-sentiment}}
 \subfloat[Liberty during COVID-19]{
 \includegraphics[width=0.49\linewidth]{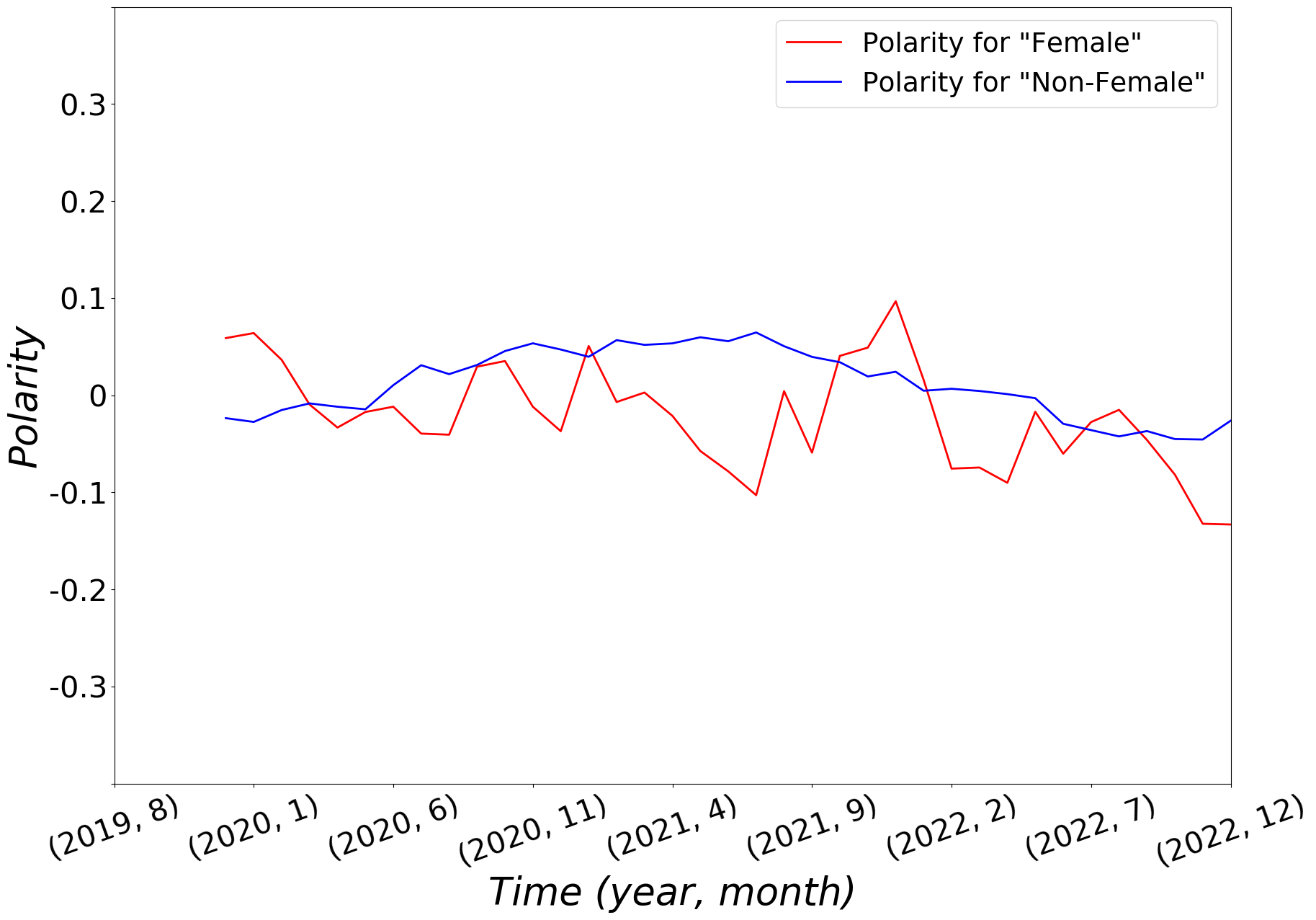}\label{fig:covid-Liberty}}
\end{figure*}

Focusing on the period of the COVID-19 pandemic, as already anticipated, we have detected a significant decrease in the extracted sentiment and moral polarities for ``Females'' with respect to the ``Non-Female''.
In Figure \ref{fig:covid_zoom_sentiment_liberty}, we depict the
\textit{sentiment} and the \textit{liberty} polarities during the COVID-19 period for the articles talking about the (un)employment and labour market topics for ``Female'' (in red) and ``Non-Female'' (in blue), using a 90 days smoothing window. While in Figure \ref{fig:sentiment_spain} we have observed an important drop in sentiment for the overall (un)employment and labour market topics, examining Figure \ref{fig:covid-sentiment} we notice that this drop in the majority of cases is related to 
a marked decrease in females (un)employment and labour market conditions. A similar behaviour, although less pronounced, is evident also for the liberty moral value.
This analysis further complements official statistics and surveys which have shown proof of the difficulties females face in the labour market in Spain \cite{Larranaga201283,RePEc:mod:dembwp:0010, Campos-Soria20164143}. 

\section{Conclusions}

In this contribution 
we have explored the press' narratives and emerging gender inequalities in the labour market analysing the Spanish press outlets over the past two decades. Extracting the sentiment and moral values reflected in the news we find that when the Spanish labour market deteriorates, like e.g. during the Great Spanish Recession, the EU Sovereign Debt Crisis, and the recent spread of the COVID-19 pandemic, news signals show a more negative trend for female unemployment.
The hardships females are facing in the job market are well-known, but capturing this phenomenon in an data-driven way directly from the press opens up diverse perspectives for further future analysis, complementing existing practices. 
Our findings will hopefully pave the way for a better understanding on how the actions of policymakers and stakeholders in Spain are reflected in the press outlets.
This analysis can be directly generalised to any linguistic sources capturing worldviews from social media and any user generated text in near real-time. 

Certainly this is an experimental work and might carry some limitations. The use of moral psychology-based lexicons is quite novel and as such the employed lexicons appear to be still limited, relative to the large dictionaries of sentiment polarities already around since a while in the field.
Although the strength of Moral Foundations Theory has been demonstrated, it still remains a controversial theory given from a lack of availability of robust lexicons to operationalize it. As such, these moral values resources should be elaborated more in depth order to be able to produce more accurate and wider results, and to devise custom analyses in diverse fields, ranging e.g. from healthcare to economics and finance. 

Furthermore, the entire pipeline rely on the English language, and therefore the quality of the translation is key to the success of the downstream NLP pipeline. Although the performance of the eTranslation service appears satisfactory to our analysis, for future works it is possible 
to customize 
the algorithm pipeline directly in the original language, avoiding in this way some unavoidable noise deriving from the translation process and producing more accurate results.

\section{Acknowledgments}
We would like to thank the colleagues of the Joint Research Centre of the European Commission for their support. The views expressed are purely those of the authors and may not in any circumstance be regarded as stating an official position of the European Commission. \\
KK gratefully acknowledges the support from the Lagrange Project of the Institute for Scientific Interchange Foundation (ISI Foundation) funded by Fondazione Cassa di Risparmio di Torino (Fondazione CRT).

\bibliography{aaai22}

\begin{thebibliography}{47}
\providecommand{\natexlab}[1]{#1}

\bibitem[{Addabbo, Rodríguez-Modroño, and
  Gálvez-Muñoz(2013)}]{RePEc:mod:dembwp:0010}
Addabbo, T.; Rodríguez-Modroño, P.; and Gálvez-Muñoz, L. 2013.
\newblock {Gender and the Great Recession: Changes in labour supply in Spain}.
\newblock Department of economics (demb), University of Modena and Reggio
  Emilia, Department of Economics "Marco Biagi".

\bibitem[{Afonso and Verdial(2020)}]{eubondSpainRep}
Afonso, A.; and Verdial, N. 2020.
\newblock {Sovereign Debt Crisis in Portugal and Spain}.
\newblock Econpol working paper 40, European Network for Economic and Fiscal
  Policy Research.

\bibitem[{Araque, Gatti, and Kalimeri(2020)}]{Araque2020}
Araque, O.; Gatti, L.; and Kalimeri, K. 2020.
\newblock MoralStrength: Exploiting a moral lexicon and embedding similarity
  for moral foundations prediction.
\newblock \emph{Knowledge-Based Systems}, 191.

\bibitem[{Araque, Gatti, and Kalimeri(2022)}]{Araque2022154}
Araque, O.; Gatti, L.; and Kalimeri, K. 2022.
\newblock LibertyMFD: A Lexicon to Assess the Moral Foundation of Liberty.
\newblock In \emph{ACM International Conference Proceeding Series}, 154 –
  160.

\bibitem[{Baccianella, Esuli, and
  Sebastiani(2010)}]{baccianella2010sentiwordnet}
Baccianella, S.; Esuli, A.; and Sebastiani, F. 2010.
\newblock {SentiWordNet 3.0: An enhanced lexical resource for sentiment
  analysis and opinion mining}.
\newblock In \emph{LREC}, volume~10, 2200--2204.

\bibitem[{Barbaglia, Consoli, and
  Manzan(2022{\natexlab{a}})}]{barbaglia2022forecasting}
Barbaglia, L.; Consoli, S.; and Manzan, S. 2022{\natexlab{a}}.
\newblock Forecasting {GDP} in {E}urope with textual data.
\newblock \emph{SSRN Working Paper. Available at:
  \url{https://dx.doi.org/10.2139/ssrn.3898680}}.

\bibitem[{Barbaglia, Consoli, and Manzan(2022{\natexlab{b}})}]{Barbaglia2022}
Barbaglia, L.; Consoli, S.; and Manzan, S. 2022{\natexlab{b}}.
\newblock Forecasting with Economic News.
\newblock \emph{Journal of Business and Economic Statistics}, to appear.

\bibitem[{Bert{\`e}, Kalimeri, and Paolotti(2023)}]{berte2023monitoring}
Bert{\`e}, M.; Kalimeri, K.; and Paolotti, D. 2023.
\newblock {Monitoring Gender Gaps via LinkedIn Advertising Estimates: The case
  study of Italy}.
\newblock \emph{arXiv preprint arXiv:2303.05862}.

\bibitem[{Blei, Ng, and Jordan(2003)}]{blei2003latent}
Blei, D.; Ng, A.; and Jordan, M. 2003.
\newblock Latent {D}irichlet {A}llocation.
\newblock \emph{Journal of Machine Learning Research}, 3: 993--1022.

\bibitem[{Bonanomi et~al.(2017)Bonanomi, Rosina, Cattuto, and
  Kalimeri}]{bonanomi2017understanding}
Bonanomi, A.; Rosina, A.; Cattuto, C.; and Kalimeri, K. 2017.
\newblock Understanding youth unemployment in Italy via social media data.
\newblock In \emph{28th IUSSP International Population Conference, Cape Town,
  South Africa}.

\bibitem[{Burnap et~al.(2016)Burnap, Gibson, Sloan, Southern, and
  Williams}]{Burnap2016230}
Burnap, P.; Gibson, R.; Sloan, L.; Southern, R.; and Williams, M. 2016.
\newblock 140 characters to victory?: Using Twitter to predict the UK 2015
  General Election.
\newblock \emph{Electoral Studies}, 41: 230 – 233.

\bibitem[{Cambria et~al.(2010)Cambria, Speer, Havasi, and
  Hussain}]{cambria2010senticnet}
Cambria, E.; Speer, R.; Havasi, C.; and Hussain, A. 2010.
\newblock {SenticNet: A publicly available semantic resource for opinion
  mining}.
\newblock In \emph{2010 AAAI Fall Symposium Series}, volume FS-10-02, 14--18.

\bibitem[{Campos-Soria and Ropero-García(2016)}]{Campos-Soria20164143}
Campos-Soria, J.~A.; and Ropero-García, M.~A. 2016.
\newblock Gender segregation and earnings differences in the Spanish labour
  market.
\newblock \emph{Applied Economics}, 48(43): 4143 – 4155.

\bibitem[{Ceron et~al.(2014)Ceron, Curini, Iacus, and Porro}]{Ceron2014340}
Ceron, A.; Curini, L.; Iacus, S.~M.; and Porro, G. 2014.
\newblock Every tweet counts? How sentiment analysis of social media can
  improve our knowledge of citizens' political preferences with an application
  to Italy and France.
\newblock \emph{New Media and Society}, 16(2): 340 – 358.

\bibitem[{Colagrossi et~al.(2022)Colagrossi, Consoli, Panella, and
  Barbaglia}]{Colagrossi2022323}
Colagrossi, M.; Consoli, S.; Panella, F.; and Barbaglia, L. 2022.
\newblock {Tracking socio-economic activities in European countries with
  unconventional data}.
\newblock In \emph{ACM International Conference Proceeding Series}, 323 –
  330.

\bibitem[{Consoli, Barbaglia, and Manzan(2020)}]{Consoli202052}
Consoli, S.; Barbaglia, L.; and Manzan, S. 2020.
\newblock Fine-grained, aspect-based semantic sentiment analysis within the
  economic and financial domains.
\newblock In \emph{Proceedings - 2020 IEEE 2nd International Conference on
  Cognitive Machine Intelligence, CogMI 2020}, 52 – 61.

\bibitem[{Consoli, Barbaglia, and Manzan(2021)}]{Consoli202187}
Consoli, S.; Barbaglia, L.; and Manzan, S. 2021.
\newblock Explaining sentiment from Lexicon.
\newblock In \emph{CEUR Workshop Proceedings}, volume 2918, 87 – 95.

\bibitem[{Consoli, Barbaglia, and Manzan(2022)}]{Consoli2022a}
Consoli, S.; Barbaglia, L.; and Manzan, S. 2022.
\newblock Fine-grained, aspect-based sentiment analysis on economic and
  financial lexicon.
\newblock \emph{Knowledge-Based Systems}, 247: 108781.

\bibitem[{Consoli, Pezzoli, and Tosetti(2021)}]{Consoli2021}
Consoli, S.; Pezzoli, L.~T.; and Tosetti, E. 2021.
\newblock Emotions in macroeconomic news and their impact on the European bond
  market.
\newblock \emph{Journal of International Money and Finance}, 118.

\bibitem[{Consoli, {Reforgiato Recupero}, and Saisana(2021)}]{Consoli20211}
Consoli, S.; {Reforgiato Recupero}, D.; and Saisana, M. 2021.
\newblock \emph{Data Science for Economics and Finance: Methodologies and
  Applications}.
\newblock Switzerland AG: Springer Nature.

\bibitem[{Consoli, Tiozzo~Pezzoli, and Tosetti(2022)}]{Consoli2022}
Consoli, S.; Tiozzo~Pezzoli, L.; and Tosetti, E. 2022.
\newblock Neural forecasting of the Italian sovereign bond market with economic
  news.
\newblock \emph{Journal of the Royal Statistical Society. Series A: Statistics
  in Society}.

\bibitem[{{Council of Europe}(2011)}]{CouncilofEuropeConvention2011}
{Council of Europe}. 2011.
\newblock The Council of Europe Convention on preventing and combating violence
  against women and domestic violence (Istanbul Convention).
\newblock
  \url{https://www.coe.int/en/web/gender-matters/council-of-europe-convention-on-preventing-and-combating-violence-against-women-and-domestic-violence}.
\newblock Accessed: 2023-04-26.

\bibitem[{Dancausa~Millán et~al.(2021)Dancausa~Millán, Millán Vázquez de~la
  Torre, Hernández~Rojas, and Jimber Del~Río}]{genderUnemployment}
Dancausa~Millán, M.~G.; Millán Vázquez de~la Torre, M.~G.; Hernández~Rojas,
  R.; and Jimber Del~Río, J.~A. 2021.
\newblock {The Spanish Labor Market: A Gender Approach}.
\newblock \emph{International Journal of Environmental Research and Public
  Health}, 18(5): 2742.

\bibitem[{Einav and Levin(2014)}]{Einav2014}
Einav, L.; and Levin, J. 2014.
\newblock Economics in the age of big data.
\newblock \emph{Science}, 346(6210): 715--721.

\bibitem[{Gast{\'o}n~Beir{\'o} et~al.(2022)Gast{\'o}n~Beir{\'o}, D'Ignazi,
  Prado, Perez~Bustos, and Kalimeri}]{gaston2022moral}
Gast{\'o}n~Beir{\'o}, M.; D'Ignazi, J.; Prado, M.~F.; Perez~Bustos, V.; and
  Kalimeri, K. 2022.
\newblock Moral Narratives Around the Vaccination Debate on Facebook.
\newblock \emph{arXiv e-prints}, arXiv--2206.

\bibitem[{Haidt and Graham(2007)}]{Haidt200798}
Haidt, J.; and Graham, J. 2007.
\newblock When morality opposes justice: Conservatives have moral intuitions
  that liberals may not recognize.
\newblock \emph{Social Justice Research}, 20(1): 98 – 116.

\bibitem[{Haidt and Joseph(2004)}]{Haidt200455}
Haidt, J.; and Joseph, C. 2004.
\newblock Intuitive ethics: how innately prepared intuitions generate
  culturally variable virtues Daedalus.
\newblock \emph{Daedalus}, 55 – 66.

\bibitem[{Hofmann et~al.(2013)Hofmann, Beverungen, Räckers, and
  Becker}]{Hofmann2013387}
Hofmann, S.; Beverungen, D.; Räckers, M.; and Becker, J. 2013.
\newblock What makes local governments' online communications successful?
  Insights from a multi-method analysis of Facebook.
\newblock \emph{Government Information Quarterly}, 30(4): 387 – 396.

\bibitem[{Iyyer et~al.(2014)Iyyer, Enns, Boyd-Graber, and
  Resnik}]{Iyyer20141113}
Iyyer, M.; Enns, P.; Boyd-Graber, J.; and Resnik, P. 2014.
\newblock Political ideology detection using recursive neural networks.
\newblock In \emph{52nd Annual Meeting of the Association for Computational
  Linguistics, ACL 2014 - Proceedings of the Conference}, volume~1, 1113 –
  1122.

\bibitem[{Jansen, Jiménez-Martín, and
  Gorjón(2016)}]{RePEc:fda:fdaeee:eee2016-10}
Jansen, M.; Jiménez-Martín, S.; and Gorjón, L. 2016.
\newblock {The Legacy of the Crisis: The Spanish Labour Market in the Aftermath
  of the Great Recession}.
\newblock Studies on the Spanish Economy eee2016-10, FEDEA.

\bibitem[{Kalimeri et~al.(2019)Kalimeri, G.~Beir{\'o}, Urbinati, Bonanomi,
  Rosina, and Cattuto}]{kalimeri2019human}
Kalimeri, K.; G.~Beir{\'o}, M.; Urbinati, A.; Bonanomi, A.; Rosina, A.; and
  Cattuto, C. 2019.
\newblock Human values and attitudes towards vaccination in social media.
\newblock In \emph{Companion Proceedings of The 2019 World Wide Web
  Conference}, 248--254.

\bibitem[{Kuypers(2020)}]{kuypers2020president}
Kuypers, J.~A. 2020.
\newblock \emph{President Trump and the news media: Moral foundations, framing,
  and the nature of press bias in America}.
\newblock Lexington Books.

\bibitem[{Lane(2012)}]{Lane2012}
Lane, P. 2012.
\newblock The European Sovereign Debt Crisis.
\newblock \emph{Journal of Economic Perspectives}, 26(49-68): 108781.

\bibitem[{Larrañaga, Valencia, and Ortiz(2012)}]{Larranaga201283}
Larrañaga, M.; Valencia, J.~F.; and Ortiz, G. 2012.
\newblock The effects of gender asymmetry in the social representation of
  female unemployment; [Efectos de la asimetría de género en la
  representación social del desempleo femenino].
\newblock \emph{Psykhe}, 21(1): 83 – 98.

\bibitem[{Lester(2001)}]{Lester2001310}
Lester, D. 2001.
\newblock Regional variations in male and female unemployment.
\newblock \emph{Perceptual and Motor Skills}, 93(PART 1): 310.

\bibitem[{Lewandowska-Gwarda(2018)}]{LewandowskaGwarda2018183}
Lewandowska-Gwarda, K. 2018.
\newblock {Female unemployment and its determinants in Poland in 2016 from the
  spatial perspective}.
\newblock \emph{Oeconomia Copernicana}, 9(2): 183 – 204.

\bibitem[{Malo(2015)}]{iloreport}
Malo, M.~A. 2015.
\newblock {Labour Market Measures in Spain 2008–13: The Crisis and Beyond}.
\newblock {Inventory of Labour Market Policy Measures in the EU 2008-13},
  International Labour Organization.

\bibitem[{Mejova, Kalimeri, and Morales(2023)}]{mejova2023authority}
Mejova, Y.; Kalimeri, K.; and Morales, G. D.~F. 2023.
\newblock Authority without Care: Moral Values behind the Mask Mandate
  Response.
\newblock \emph{arXiv preprint arXiv:2303.12014}.

\bibitem[{Mikolov et~al.(2013)Mikolov, Sutskever, Chen, Corrado, and
  Dean}]{Mikolov2013}
Mikolov, T.; Sutskever, I.; Chen, K.; Corrado, G.; and Dean, J. 2013.
\newblock Distributed representations of words and phrases and their
  compositionality.
\newblock In \emph{Advances in Neural Information Processing Systems (NIPS
  2013)}, 3111–3119. United States: ACM.

\bibitem[{Mohammad et~al.(2015)Mohammad, Zhu, Kiritchenko, and
  Martin}]{Mohammad2015480}
Mohammad, S.~M.; Zhu, X.; Kiritchenko, S.; and Martin, J. 2015.
\newblock Sentiment, emotion, purpose, and style in electoral tweets.
\newblock \emph{Information Processing and Management}, 51(4): 480 – 499.

\bibitem[{Neuhäuser(2011)}]{Neuhauser20111}
Neuhäuser, M. 2011.
\newblock \emph{{Nonparametric statistical tests: A computational approach}}.
\newblock Boca Raton, US: Chapman \& Hall.

\bibitem[{Rama et~al.(2020)Rama, Mejova, Tizzoni, Kalimeri, and
  Weber}]{rama2020facebook}
Rama, D.; Mejova, Y.; Tizzoni, M.; Kalimeri, K.; and Weber, I. 2020.
\newblock Facebook ads as a demographic tool to measure the urban-rural divide.
\newblock In \emph{Proceedings of The Web Conference 2020}, 327--338.

\bibitem[{Tatli and Tasci(2021)}]{Tatli2021225}
Tatli, H.; and Tasci, K. 2021.
\newblock {The short and long-term relation between human development and
  female unemployment: The case of Turkey}.
\newblock \emph{Argumenta Oeconomica}, 47(2): 225 – 252.

\bibitem[{Tumasjan et~al.(2011)Tumasjan, Sprenger, Sandner, and
  Welpe}]{Tumasjan2011402}
Tumasjan, A.; Sprenger, T.~O.; Sandner, P.~G.; and Welpe, I.~M. 2011.
\newblock Election forecasts with Twitter: How 140 characters reflect the
  political landscape.
\newblock \emph{Social Science Computer Review}, 29(4): 402 – 418.

\bibitem[{Urbinati et~al.(2020)Urbinati, Kalimeri, Bonanomi, Rosina, Cattuto,
  and Paolotti}]{urbinati2020young}
Urbinati, A.; Kalimeri, K.; Bonanomi, A.; Rosina, A.; Cattuto, C.; and
  Paolotti, D. 2020.
\newblock Young adult unemployment through the lens of social media: Italy as a
  case study.
\newblock In \emph{Social Informatics: 12th International Conference, SocInfo
  2020, Pisa, Italy, October 6--9, 2020, Proceedings 12}, 380--396. Springer.

\bibitem[{Xu et~al.(2012)Xu, Jun, Zhu, and Bellmore}]{Xu2012656}
Xu, J.-M.; Jun, K.-S.; Zhu, X.; and Bellmore, A. 2012.
\newblock Learning from bullying traces in social media.
\newblock In \emph{NAACL HLT 2012 - 2012 Conference of the North American
  Chapter of the Association for Computational Linguistics: Human Language
  Technologies, Proceedings of the Conference}, 656 – 666.

\bibitem[{Zolnik(2013)}]{Zolnik201376}
Zolnik, E.~J. 2013.
\newblock {A spatial analysis of male and female unemployment in the USA}.
\newblock \emph{International Journal of Applied Geospatial Research}, 4(4): 76
  – 87.

\end{thebibliography}

\end{document}